\documentclass[preprint,12pt,number]{elsarticle}

\usepackage{amssymb}
\usepackage{amsmath}
\usepackage[normalem]{ulem}
\usepackage{soul}
\usepackage{xcolor}
\usepackage{caption}
\usepackage{hyperref}

\usepackage{hyperref}
\newcounter{aqctr}
\newenvironment{author-query}
{\refstepcounter{aqctr}\par\vspace{\baselineskip}\noindent
\color{red}\textbf{Author Query/Comment AQ \arabic{aqctr}.}}
{\par\vspace{\baselineskip}\normalcolor}

\journal{Nuclear Physics B}

\begin{document}

\begin{frontmatter}



\title{Penetration of impact-induced jets into skin-simulating materials } 


\author[a]{Kohei Yamagata}
\author[b]{Yuto Yokoyama}
\author[a]{Shoto Sekiguchi}
\author[a]{Hiroya Watanabe}
\author[a,c]{Prasad Sonar}
\author[a]{Yoshiyuki Tagawa\fnref{corstar}}

\fntext[corstar]{Corresponding author: Yoshiyuki Tagawa. Email: \texttt{tagawayo@cc.tuat.ac.jp}}

\affiliation[a]{organization={Department of Mechanical Systems Engineering, Tokyo University of Agriculture and Technology},
            addressline={Nakacho 2-24-16}, 
            city={Koganei},
            postcode={184-8588}, 
            state={Tokyo},
            country={Japan}}

\affiliation[b]{organization={Micro/Bio/Nanofluidics Unit, Okinawa Institute of Science and Technology},
            addressline={Onna-son 1919-1}, 
            city={Kunigami-gun},
            postcode={904-0497}, 
            state={Okinawa},
            country={Japan}}

\affiliation[c]{
    Department and Organization={Present address: Central Food Technological Research Institute}, 
    addressline={Cheluvamba Mansion, Mysore}, 
    city={Mysore}, 
    postcode={570020}, 
    state={Karnataka}, 
    country={India}
}
\begin{abstract}
This study compares the penetration characteristics of impact-induced jets with those of laser-induced jets, focusing on the underlying penetration mechanism rather than device performance for needle-free injection. Using an impact-induced jet system capable of ejecting a highly focused liquid jet at high speed without the use of lasers, we examine jet penetration into skin-simulating materials. Unlike conventional needle-free injectors that produce diffused liquid jets, the impact-induced method generates a highly focused jet that limits the injected area, thereby reducing invasiveness. Comparative experiments with laser-induced jets show that, even at similar jet tip velocities, impact-induced jets achieve greater penetration depth. The penetration depth remains constant regardless of the offset distance $D$ from the target, owing to the high and nearly uniform velocity of the cylindrical jet root region, indicating that penetration is governed by the cylindrical jet structure. Furthermore, we systematically vary the liquid viscosity, jet inertia, and elastic modulus of the skin-simulating material. To account for cylindrical liquid jet penetration, a shear deformation model is proposed, in which the jet kinetic energy is dissipated through deformation of the material. The model shows good agreement with experimental results and provides a unified physical basis for liquid jet penetration.
\end{abstract}

\begin{graphicalabstract}
\centering
\includegraphics[width=1\textwidth]{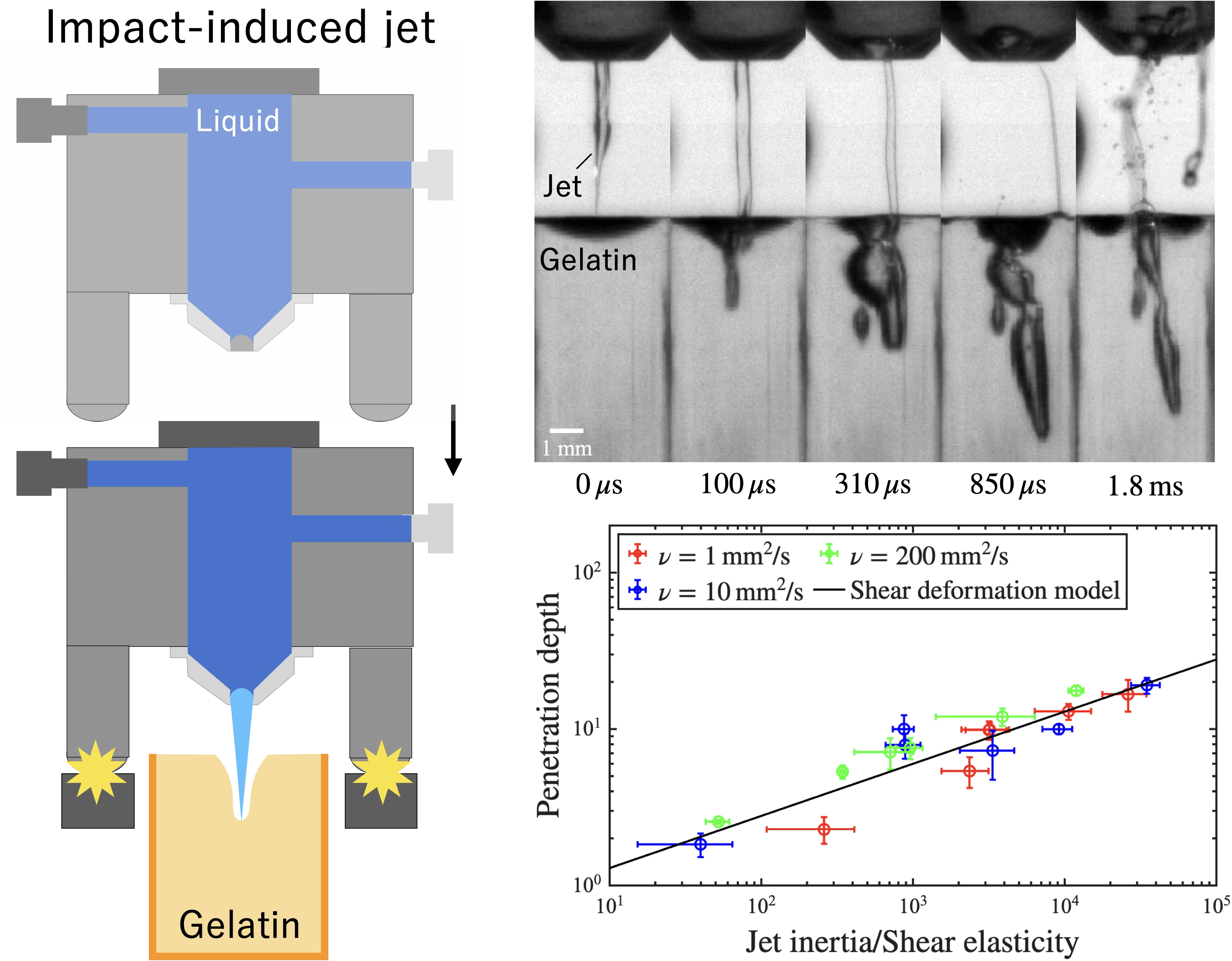}
\end{graphicalabstract}

\begin{highlights}
\item The penetration depth of impact-induced jets is governed by the high velocity of the jet root, rather than by the jet tip velocity.
\item Compared with laser-induced jets at comparable jet tip velocities, impact-induced jets exhibit significantly deeper penetration due to their higher jet root velocities.
\item The penetration depth of impact-induced jets is independent of the offset distance $D$ from the target, because the cylindrical jet root region maintains nearly uniform velocity regardless of the distance $D$.
\item Systematic experiments varying the liquid viscosity, jet velocity, and material stiffness reveal clear deviations from the conventional penetration model.
\item To address this limitation, a new shear deformation model is proposed, demonstrating that penetration is governed by shear deformation of the skin-simulating material.
\item The shear deformation model provides a unified physical framework for liquid jet penetration and offers guidance for future needle-free drug delivery applications.
\end{highlights}


\end{frontmatter}



\section{Introduction}
Needle-free injection has attracted attention as a method of avoiding needlestick injuries, disposal problems, and pain or fear associated with needles~\citep{Freeman2021,Garg,kumar,taberner2012needle,miyazaki2019development,mckeage2017high}.

\begin{figure}[htbp]
    \begin{center}
        \includegraphics[width=1\columnwidth]{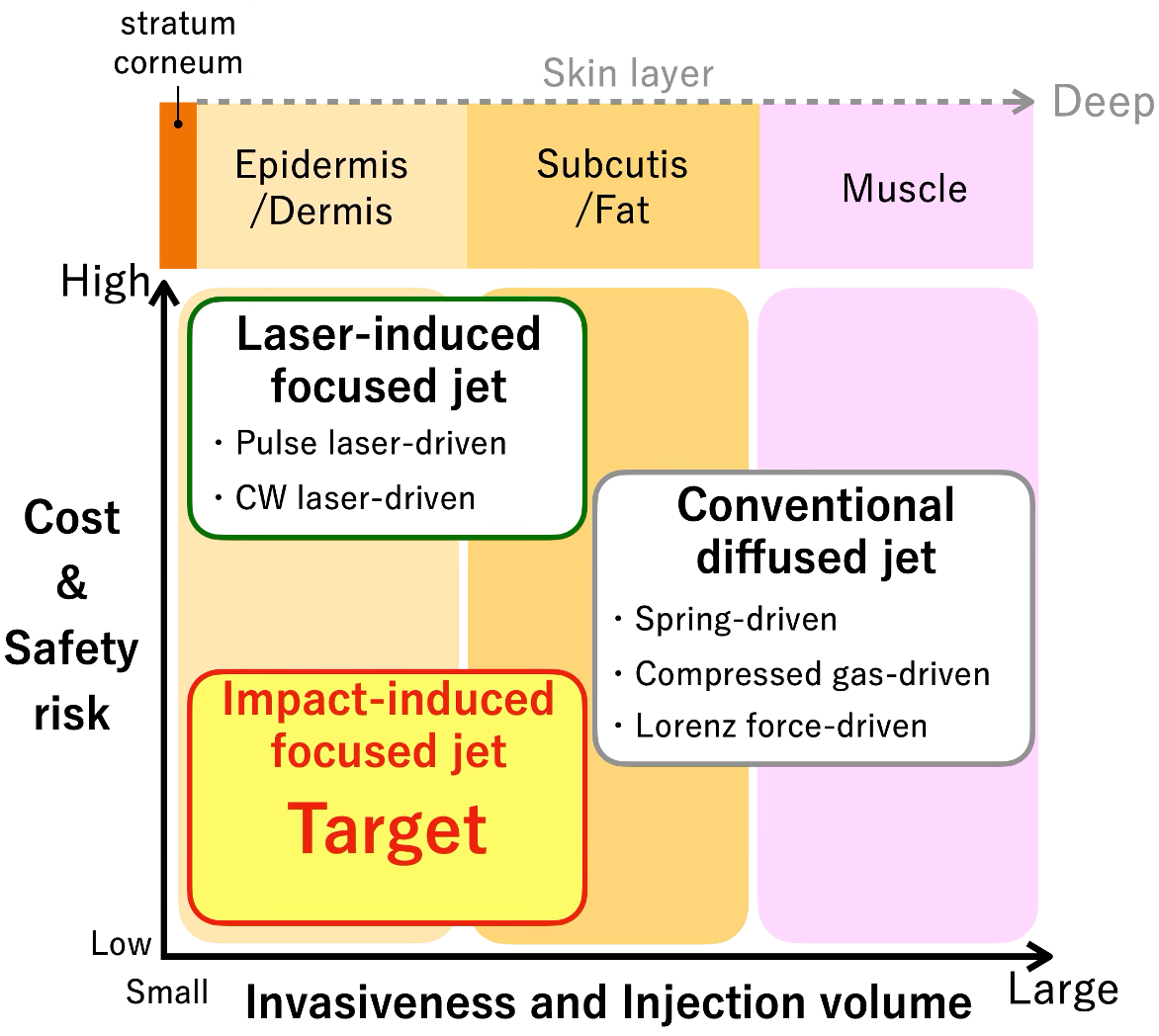}
        \caption{A positioning map comparing the cost, safety risk, invasiveness, and injection volume of a conventional needle-free injector, a laser-induced jet, and an impact-induced jet.
The vertical axis represents cost and safety risk, and the horizontal axis represents invasiveness and injection volume.
}
        \label{fig:/Positioning_map_fig}
    \end{center}
\end{figure}

A comparison of conventional needle-free injectors and laser-induced jets is illustrated in Fig.~\ref{fig:/Positioning_map_fig} and highlights the trade-offs between cost, safety, invasiveness, and injection volume. Various needle-free injection (NFI) technologies have been developed using spring-driven~\citep{taberner2012needle,rohilla2020loading}, compressed gas-driven~\citep{miyazaki2019development,mohizin2020effect}, and lorentz force-driven systems~\citep{mckeage2017high,mckeage2018power}. As shown in Fig.~\ref{fig:/Positioning_map_fig}, these conventional devices mainly target deeper injections, such as those into the muscle layer, and commonly produce diffused jets that expand wider than the nozzle orifice, causing excessive injection volume and strong stimulation of pain receptors~\citep{schoppink2022jet,miyazaki2021dynamic,rodriguez2017toward,krizek2020needle}. To address these issues, laser-induced jets have been proposed as a new generation of focused jet injection systems~\citep{tagawa2012highly,rohilla2023focused,schoppink2024laser,krizek2020repetitive,delrot2018depth}. Their high-speed and highly focused jets are well suited for shallow skin delivery, where they can target langerhans cells located near the epidermal surface, enabling strong immune responses with a small dose~\citep{schoppink2022jet,rentzsch2021specific,chen2002targeting,neagu2022langerhans}. Because such jets are thinner than the nozzle diameters, they enable precise small-volume drug delivery~\citep{schoppink2022jet,tagawa2013needle}, achieving accurate dose control with minimal pain and reduced dispersion compared to diffused jets~\citep{schoppink2022jet,krizek2020needle,tagawa2013needle,rodriguez2017toward}. However, the use of lasers introduces major challenges, including cavitation damage~\citep{sreedhar2017cavitation,fu2024secondary}, drug thermal denaturation~\citep{schoppink2022jet,lukavc2019variable}, and high device costs~\citep{oyarte2020microfluidics}. 

\begin{figure}[htbp]
    \begin{center}
        \includegraphics[width=1\columnwidth]{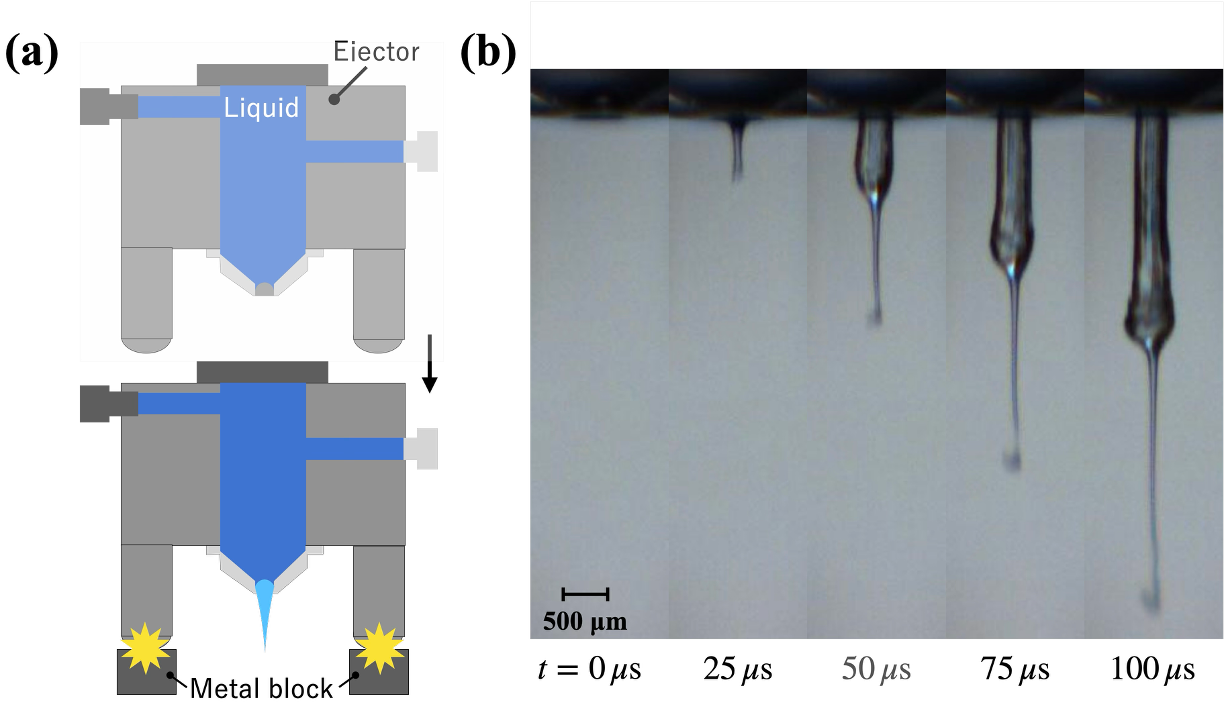}
        \caption{(a) A sketch of an impact-induced jet. The ejector is dropped downward to collide with a metal block, generating a high-speed jet upon impact. (b) A high-speed image sequence of a focused impact-induced jet ejected with silicone oil (kinematic viscosity: $10~\mathrm{mm^2/s}$).
}
        \label{fig:/Impact_induced_jet_fig}
    \end{center}
\end{figure}

Here, we focus on a new focused jet generation technique, impact-induced jet generation, as a cost-effective, laser-free alternative. This mechanism produces a focused jet by accelerating liquid through impact and focusing it at a concave gas–liquid interface when the container is abruptly stopped~\citep{antkowiak2007short,kiyama2016effects,cheng2024viscous,watanabe2025effect}. Although previous designs could eject high-viscosity liquids~\citep{onuki2018microjet,kamamoto2021drop}, they did not reach the $75~\mathrm{m/s}$ velocity required for human skin penetration~\citep{tagawa2013needle}. By adopting a tapered nozzle shape, we have developed an impact-induced method for generating focused jets that exceed $100~\mathrm{m/s}$, which is sufficient to penetrate human skin~\citep{tagawa2013needle,arora2007needle,romgens2016penetration}, even for liquids with viscosities as high as $200~\mathrm{mm^2/s}$. As shown in Fig.~\ref{fig:/Impact_induced_jet_fig}, this compact and low-cost system generates focused jets without lasers and can target the “Target” region in Fig.~\ref{fig:/Positioning_map_fig}, resulting in low-cost, minimally invasive, and small-volume injection applications.

The penetration of liquid jets into skin-simulating materials has been widely studied~\citep{miyazaki2021dynamic,rodriguez2017toward,tagawa2013needle,van2023microfluidic,anderson2017analytical,baxter2005jet,ohl2025penetration,schramm2004needle, igarashi2024optimal, gonzalez2023bubble, mousavi2025modelling}. Previous studies have revealed that the penetration depth is not only determined by the representative velocity of the jet, but also by the morphology of the jet and its internal velocity distribution~\citep{miyazaki2021dynamic, rodriguez2017toward, igarashi2024optimal, gonzalez2023bubble}. 
In previous studies, the penetration behavior of liquid jets into skin-simulating materials has been investigated mainly by varying the jet inertia and the material elasticity, and the results have often been organized using the elastic Froude number $Fr_e$ ($Fr_e =$ inertial force / material elasticity)~\citep{baxter2005jet, van2023microfluidic, ohl2025penetration, kiyama2019gelatine}. 
It has also been suggested that the viscous force of the liquid plays an important role in the penetration process and that the penetration behavior can be described in terms of the Reynolds number $Re$ ($Re =$ inertial force / viscous force)~\citep{tagawa2013needle, mousavi2025modelling}. 
However, due to the technical difficulty of ejecting high-viscosity liquids at sufficiently high speeds, no systematic investigation covering a wide range of liquid viscosities has been conducted yet~\citep{mckeage2024effect, williams2019viscous, rohilla2019characterization}.

In this study, the penetration behavior of impact-induced jets capable of ejecting high-viscosity liquids into skin-simulating materials is investigated. Analysis of the internal velocity distribution of the liquid jets reveals a distinct penetration mechanism in the impact-induced jet. In this case, the penetration depth is not governed by the jet tip, as commonly assumed, but by the flow in the root region of the jet. Furthermore, systematic penetration experiments are performed by varying the inertial, viscous, and elastic forces of the liquid--solid system. On the basis of these experiments, a new penetration model, termed the shear deformation model, is proposed, in which the kinetic energy of the liquid jet is dissipated through the shear deformation of the skin-simulating materials, and the validity of this model is experimentally verified. 
We systematically vary the jet inertia, the liquid viscosity (by two orders of magnitude, $1-200~\mathrm{mm^2/s}$), and the elasticity of skin-simulating materials to establish a broadly applicable penetration model for needle-free injection applications.

\label{sec1}

\section{Materials and methodology}
We first conducted experiments to investigate differences in the penetration behavior of focused liquid jets penetrating skin-simulating materials when generated by impact-induced jets and laser-induced jets. Subsequently, we systematically varied the jet velocity of the impact-induced jets, the viscosity of the liquid, and the elasticity of the skin-simulating materials to examine their effects on the penetration dynamics.
\begin{figure}[htbp]
    \begin{center}
        \includegraphics[width=1\columnwidth]{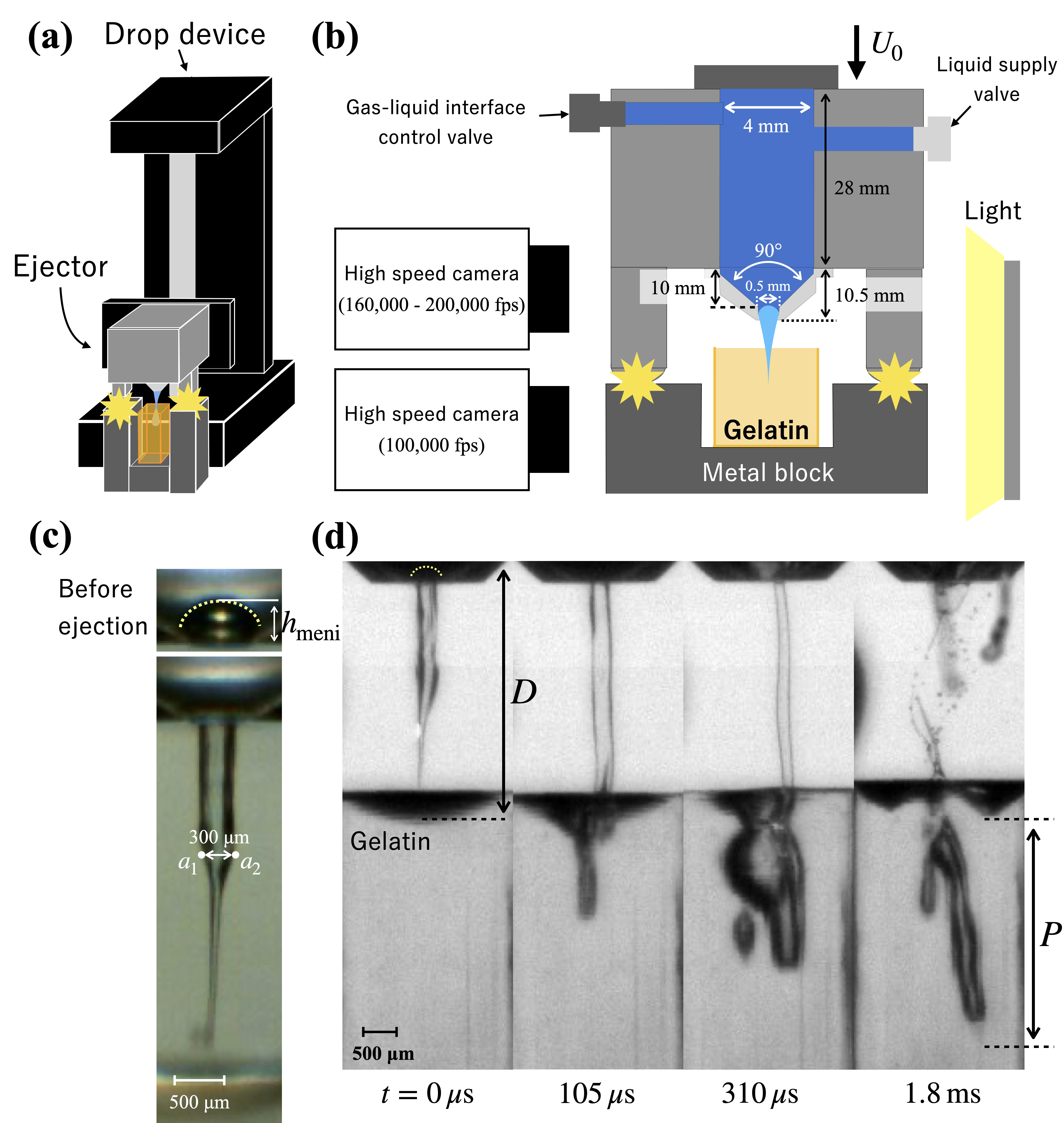}
        \caption{(a) A three-dimensional schematic of the experimental setup for the impact-induced jet. (b) A two-dimensional side-view schematic of the experimental setup. The upper high-speed camera (160{,}000-200{,}000 fps) captures the jet ejection process, while the lower high-speed camera (100{,}000 fps) captures the penetration behavior into the gelatin. (c) The gas–liquid interface before jet ejection (yellow dotted line) and a jet ejection image; $a_1$ and $a_2$ denote the two reference points located $150~\mu$m away from the nozzle-orifice center. (d) $D$ is defined as the distance from the top of the meniscus to the gelatin surface, and the penetration depth is $P$.
}
        \label{fig:Experimental_setup_fig}
    \end{center}
\end{figure}

\begin{figure}[htbp]
    \begin{center}
        \includegraphics[width=1\columnwidth]{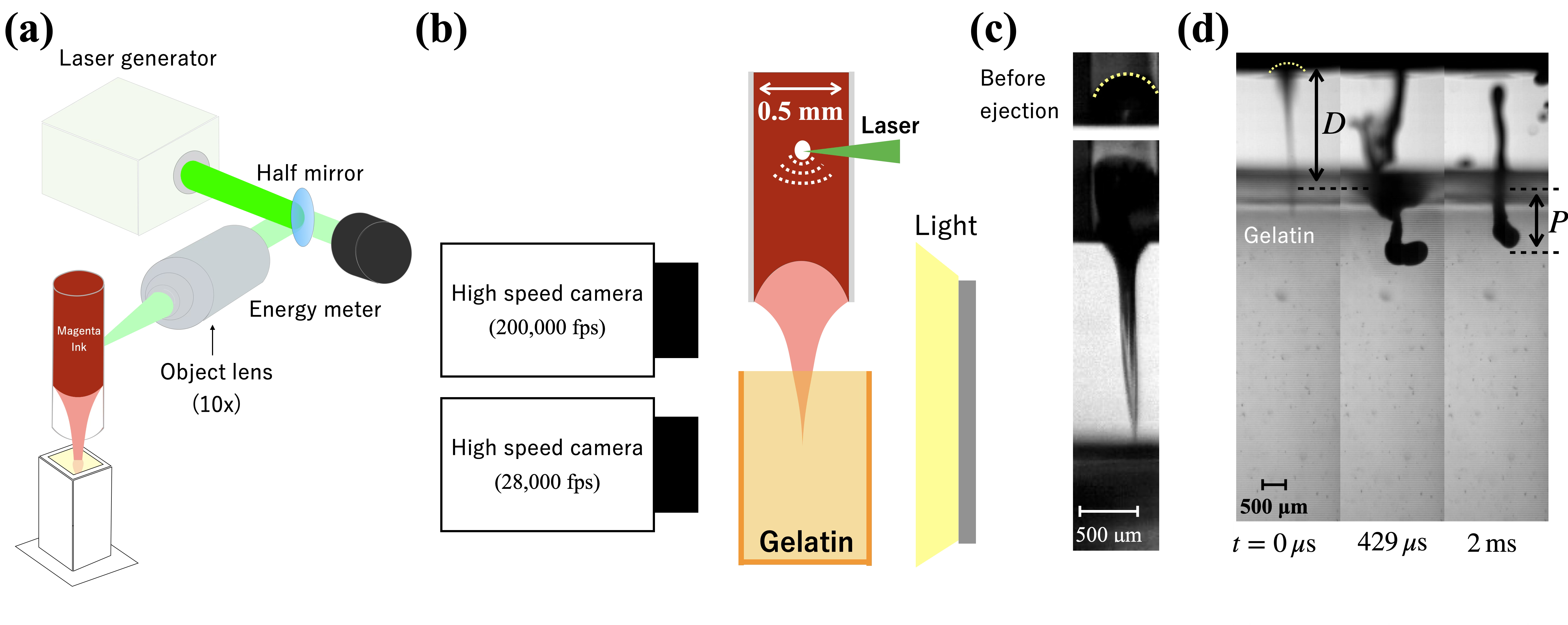}
        \caption{(a) A three-dimensional schematic of the experimental setup for the laser-induced jet. (b) A schematic of the experimental setup. The upper high-speed camera (200{,}000 fps) captures the jet ejection process, while the lower high-speed camera (28{,}000 fps) captures the penetration into the gelatin. (c) The gas–liquid interface before jet ejection (yellow dotted line) and a jet ejection image. (d) The distance $D$ is defined from the top of the meniscus to the gelatin surface and the penetration depth is $P$.
}
        \label{fig:Experimental_setup_2_fig}
    \end{center}
\end{figure}

The experimental setup with an impact-induced jet is shown in Fig.~\ref{fig:Experimental_setup_fig}(a)(b).
The ejector is dropped onto a metal block, applying an impact force and generating a focused liquid jet~\citep{kamamoto2021drop}. The jet is ejected through a focusing effect that occurs at the gas--liquid interface inside the translucent resin nozzle made of polyvinyl chloride (PVC)~\citep{antkowiak2007short}. In the experiments on impact-induced jets, 10~cSt silicone oil with a kinematic viscosity of 10 $\rm{mm^2/s}$~(KF-96 series, Shin-Etsu Chemical; silicone oil, Sigma
Aldrich) was used. The high-speed camera (FASTCAM SA-Z, Photron, Japan, temporal resolution: 160,000--200,000 frames per second) captures the jet ejection phenomena in Fig.~\ref{fig:Experimental_setup_fig}(c), and another high-speed camera (FASTCAM SA-X, Photron, Japan, temporal resolution: 100,000 frames per second) captures the penetration phenomena into the skin-simulating materials in Fig.~\ref{fig:Experimental_setup_fig}(d). 
In experiments with a laser-induced jet, as shown in Fig.~\ref{fig:Experimental_setup_2_fig}(a)(b), a capillary tube is connected to a syringe and filled with magenta ink (THC-7M4N, Elecom Co., Japan, viscosity: {2.7}~{${\rm{mPa{\cdot}s}}$}, density: {1060}~${\rm{kg/m{^3}}}$, surface tension: {40}~$\rm{mN/m}$), which has high energy-absorption efficiency for green light. A high-speed focused liquid jet is ejected by focusing a pulsed laser (wavelength: {532}~{$\rm{nm}$}, pulse width: {6}~{$\rm{ns}$}) through an objective lens (MPLN10x, Olympus Co., Japan, magnification: $10\times$, N.A. value: 0.25) onto the liquid in the tube. By forming a concave meniscus as shown in Fig.~\ref{fig:Experimental_setup_2_fig}(c), we generated a focused microjet~\citep{tagawa2012highly,peters2013highly}. A half-mirror (OptoSigma Co., transmission: 50\%) splits the pulsed laser into light directed toward the circular tube and light directed toward an energy meter (EnergyMax-RS J-10MB-HE, Coherent Co., USA) to measure the laser energy $E$. The energy of the laser in this experiment is 2.1 mJ, and the capillary tube is irradiated with laser energy of 1.05 mJ to eject the jet, which has almost the same jet tip velocity as the impact-induced jets. A high-speed camera (FASTCAM SA-X, Photron, Japan, temporal resolution: 200,000 frames per second) is used to capture the jet ejection phenomena. The lower high-speed camera (CRYSTA PI-1P, Photron, Japan; temporal resolution: 28,000 frames per second), equipped with a polarization filter array, captures the penetration into the skin-simulating materials. The two high-speed cameras and a laser generator (Nd:YAG laser; Nano S PIV, Litron Laser Co., USA) are synchronized by a delay generator (Model 575 Pulse/Delay Generator, BNC Co., USA).

Gelatin was used as a skin-simulating material. The gelatin was prepared by stirring at 70 $^\circ\mathrm{C}$ and 600 rpm for 30 minutes. Subsequently, it was placed in a highly transparent resin container fabricated using a 3D printer (Form 3, Formlabs Inc., USA, resin: clear) in the experiment with impact-induced jets, and a transparent plastic container in the experiment with laser-induced jets. The gelatin-filled containers were stored in a refrigerator at a temperature of 4 $^\circ\mathrm{C}$ for 24 hours. The gelatin concentration was adjusted to 3 and $5\ \text{wt}\%$ to replicate the required stiffness, where the  $3\ \text{wt}\%$ gelatin condition was chosen to mimic the stiffness of the human subcutis~\citep{feng2022vivo} and the $5\ \text{wt}\%$ gelatin condition was chosen to mimic the stiffness of human epidermis~\citep{tagawa2013needle,menezes2009shock}.
The motion of the injected liquid was considered stabilized 1.8~ms after the jet tip made contact with the gelatin surface, at which time the penetration process had completed. Accordingly, the penetration depth $P$ was measured from images captured 1.8~ms after the initial contact between the liquid jet tip and the gelatin surface. The distance from the top of the meniscus inside the nozzle to the surface of the gelatin was defined as $D$, as shown in Fig.~\ref{fig:Experimental_setup_fig}(d). The experiments were conducted under eight different $D$ conditions, with 4–6 trials for each.
The velocities of the jet tip and root region ($V_{\mathrm{jet\,tip}}$ and $V_{\mathrm{jet\,root}}$) were influenced by the shape of the gas–liquid interface~\citep{tagawa2012highly, antkowiak2007short}, as indicated by the yellow dotted line in Fig.~\ref{fig:Experimental_setup_fig}(c). Here, $V_{\mathrm{jet\,tip}}$ was defined as the velocity calculated from the travel distance of the jet tip from the nozzle exit to the gelatin surface, while $V_{\mathrm{jet\,root}}$ was defined as the mean velocity of the jet root region, evaluated at two points, $a_1$ and $a_2$, located $150~\mu$m from the nozzle-orifice center shown in Fig.~\ref{fig:Experimental_setup_fig}(c). In the comparative experiment between impact-induced jets and laser-induced jets, the height of the gas–liquid interface, $h_{\rm{meni}}$, was adjusted to $0.52$–$0.67~\rm{mm}$ using the gas–liquid interface control valve shown in Fig.~\ref{fig:Experimental_setup_fig}(c). 

In further experiments, the physical properties of the liquid jets and the shear modulus of gelatin were systematically varied to model the penetration behavior with impact-induced jets. Table~\ref{tab:fluid_properties} summarizes the physical properties of the liquid jets used in this study, together with the shear modulus of gelatin, for which we adopted the values of the shear modulus reported in a previous study~\citep{kiyama2019gelatine}. The experimental parameters included the diameter of the nozzle $d_{\mathrm{jet}}$, the height of the meniscus $h_{\mathrm{meni}}$ to control the shape of the liquid jet, the impact speed $U_{\mathrm{0}}$ to control the velocity of the liquid jet, the liquid density $\rho$, the kinematic viscosity $\nu$ of the liquid, and the shear modulus $G$ of the gelatin. For each experimental condition, the jet velocity ($V_{\mathrm{jet\,tip}}$ and $V_{\mathrm{jet\,root}}$), the shear modulus of gelatin, and the viscosity of the liquid were systematically varied, and 3--5 trials were conducted to ensure reproducibility. 

\begin{table}[htbp]
\centering
\caption{Experimental conditions for the liquid jet and gelatin properties.}
\label{tab:fluid_properties}
\begin{tabular}{cc}
\hline
\textbf{Parameter} & \textbf{Experimental value} \\
\hline
Nozzle diameter $d_{\mathrm{jet}}$ [mm] & 0.5 \\
Liquid density $\rho$ [kg/m$^3$] & 818, 935, 970  \\
Liquid surface tension $\sigma$ [mN/m] & 16.9, 20.1, 21.1 \\
Liquid kinematic viscosity $\nu$ [mm$^2$/s] & 1, 10, 200 \\
Meniscus height $h_{\rm{meni}}$ [mm] & 0.45--0.82 \\
Impact speed $U_{\mathrm{0}}$ [m/s] & 0.97, 1.5, 2.3 \\
Jet tip velocity $V_{\mathrm{jet\,tip}}$ [m/s] & 36 -- 116 \\
Jet root velocity $V_{\mathrm{jet\,root}}$ [m/s] & 24 -- 66 \\
Shear modulus of gelatin $G$ [Pa] & 978, 5542 \\
\hline
\end{tabular}
\end{table}

\label{sec2}

\section{Penetration behavior of the impact-induced jets and laser-induced jets}
\subsection{Experimental result}
\begin{figure}[htpb]
    \begin{center}
        \includegraphics[width=1\columnwidth]{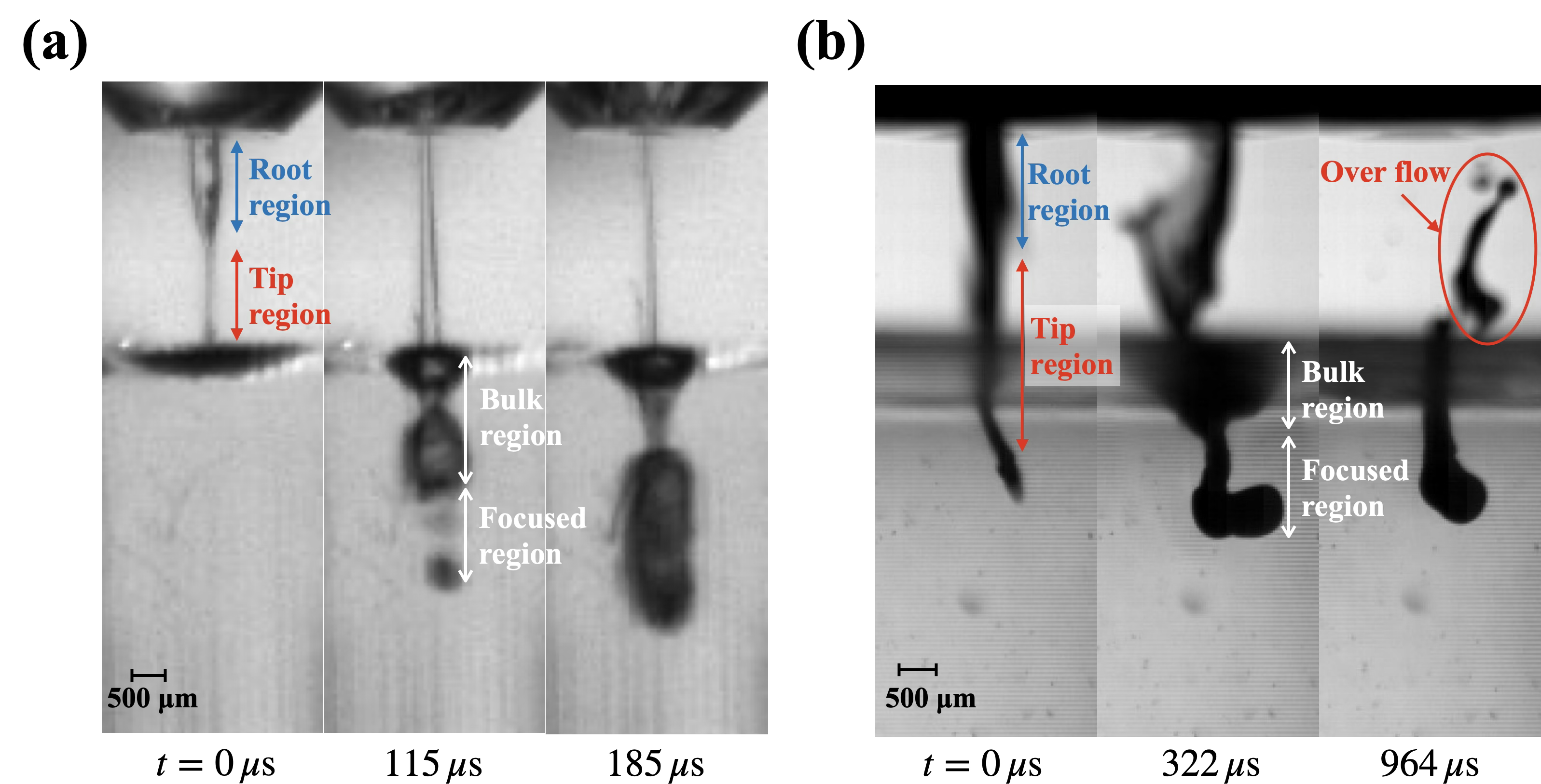}
        \caption{An image sequence of the injection of (a) an impact-induced jet (jet tip velocity $V_{\mathrm{jet\,tip}} = 88.8~\mathrm{m/s}$) and (b) a laser-induced jet (jet tip velocity $V_{\mathrm{jet\,tip}} = 86.8~\mathrm{m/s}$), using a gelatin target with a shear modulus of $5542~\mathrm{Pa}$. 
}
        \label{fig:Penetration_behavior_image_1}
    \end{center}
\end{figure}

Figure~\ref{fig:Penetration_behavior_image_1}(a)(b) shows the experimental post-penetration images obtained with the impact-induced jets and the laser-induced jets. Both had almost the same jet tip velocity, while the distance $D$ between the top of the meniscus and the gelatin surface was varied.
\begin{figure}[htpb]
    \begin{center}
        \includegraphics[width=1\columnwidth]{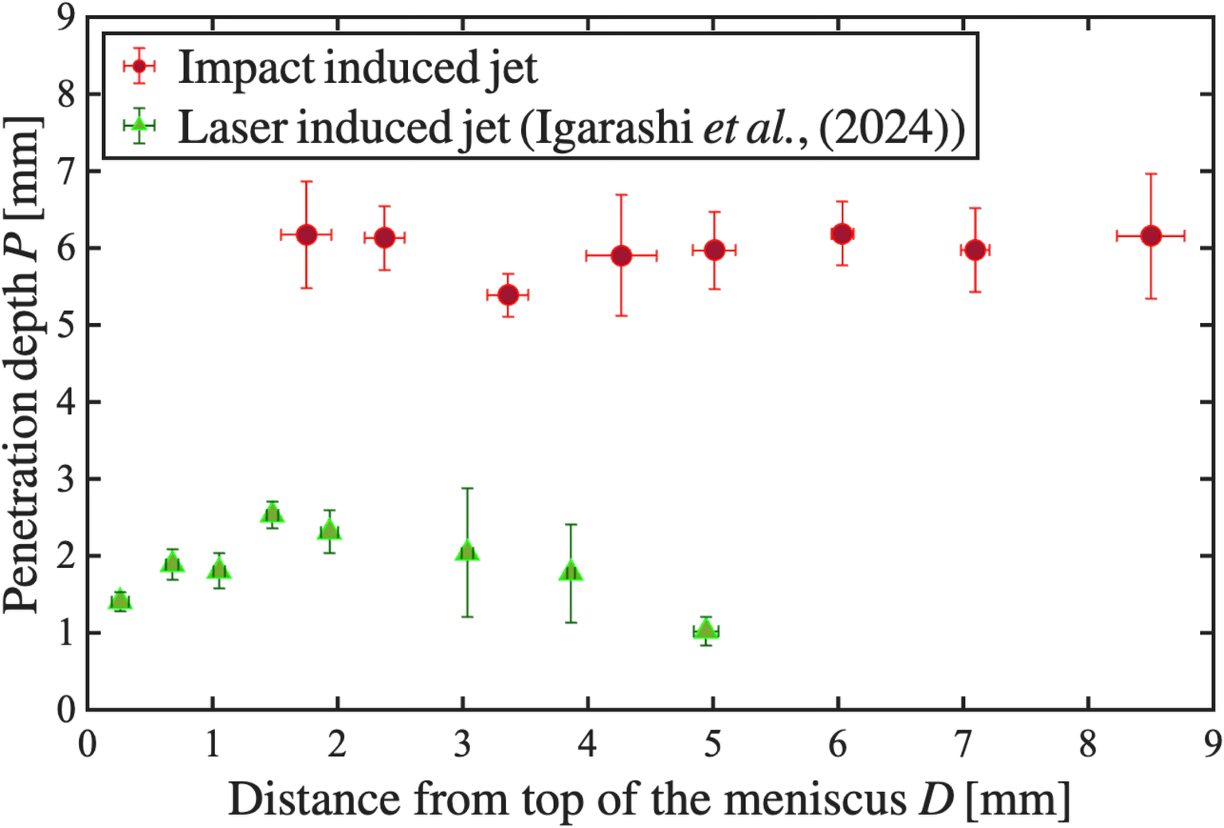}
        \caption{The penetration depth \(P\) as a function of distance \(D\) for the
  (i) impact-induced jet (\(V_{\mathrm{jet\,tip}} = 86.8 \pm 11.2~\mathrm{m/s}\))
  and the (ii) laser-induced jet\,\citep{igarashi2024optimal} (\(V_{\mathrm{jet\,tip}} = 126.9 \pm 13.8~\mathrm{m/s}\)).
  Error bars denote ±1 standard deviation.
}
        \label{fig:Penetration_depth}
    \end{center}
\end{figure}
In both cases, the jet could be divided into two characteristic regions: the tip region (indicated by red arrows), which represents the leading edge of the jet, and the root region (indicated by blue arrows), which corresponds to the backward section of the jet near the nozzle exit. Here, the root region is defined as the region in which the diameter of the jet is greater than 300~$\mu\mathrm{m}$. From these experimental results in Fig.~\ref{fig:Penetration_behavior_image_1}(a)(b), it was confirmed that the penetration depth of the impact-induced jet was much larger than that of the laser-induced jet. This difference can be attributed to the distinct penetration behaviors of the two jet generation mechanisms. As shown in Fig.~\ref{fig:Penetration_behavior_image_1}(a)(b), both jets first penetrated the gelatin with the tip region, forming a narrow penetration region called the focused region, followed by the root region. In the case of the impact-induced jet, the bulk region (originating from the root region) reached the focused region and further penetrated the gelatin, resulting in a greater penetration depth. In contrast, for the laser-induced jet, the bulk region did not penetrate but was instead expelled into the air above the gelatin, with only the focused region entering the gelatin. In the case of laser-induced jets, the velocity at the tip of the jet has been reported to be significantly higher than at the root, and the penetration depth is primarily determined by the velocity at the tip of the jet \citep{miyazaki2021dynamic,tagawa2013needle,igarashi2024optimal}. In contrast, for impact-induced jets, our experiments indicate that the root region of the jet in Fig.~\ref{fig:Penetration_behavior_image_1}(a) contributes substantially to the penetration depth. 

Figure~\ref{fig:Penetration_depth} shows the results of the $5~\mathrm{wt}\%$ gelatin experiments for the impact-induced jet, together with data for the laser-induced jet from a previous study~\citep{igarashi2024optimal}. As shown in Fig.~\ref{fig:Penetration_depth}, although the tip velocity of the impact-induced jet is smaller than that of the laser-induced jet, the measured penetration depth reaches approximately 6~mm, which is approximately twice the maximum penetration depth observed for the laser-induced jet. Moreover, in the impact-induced jet case, the penetration depth remains constant regardless of the distance~$D$, and the behavior reported in a previous study \citep{igarashi2024optimal}, where the penetration depth varied with the distance $D$ due to changes in the shape of the jet tip, is not observed. This suggests that penetration behavior is governed by the root region in impact-induced jets. In the following section, we explain why the root region of the impact-induced jet contributes significantly to the penetration depth, with a particular focus on the internal velocity distribution of the liquid jet.
\subsection{Velocity distribution within the impact-induced jet} 
In this section, we analyze and discuss the phenomenon described in the previous section, in which the root region of the impact-induced jet contributes to penetration, focusing on the velocity distribution within the impact-induced jet. 

\begin{figure}[htbp]
    \begin{center}
        \includegraphics[width=1\columnwidth]{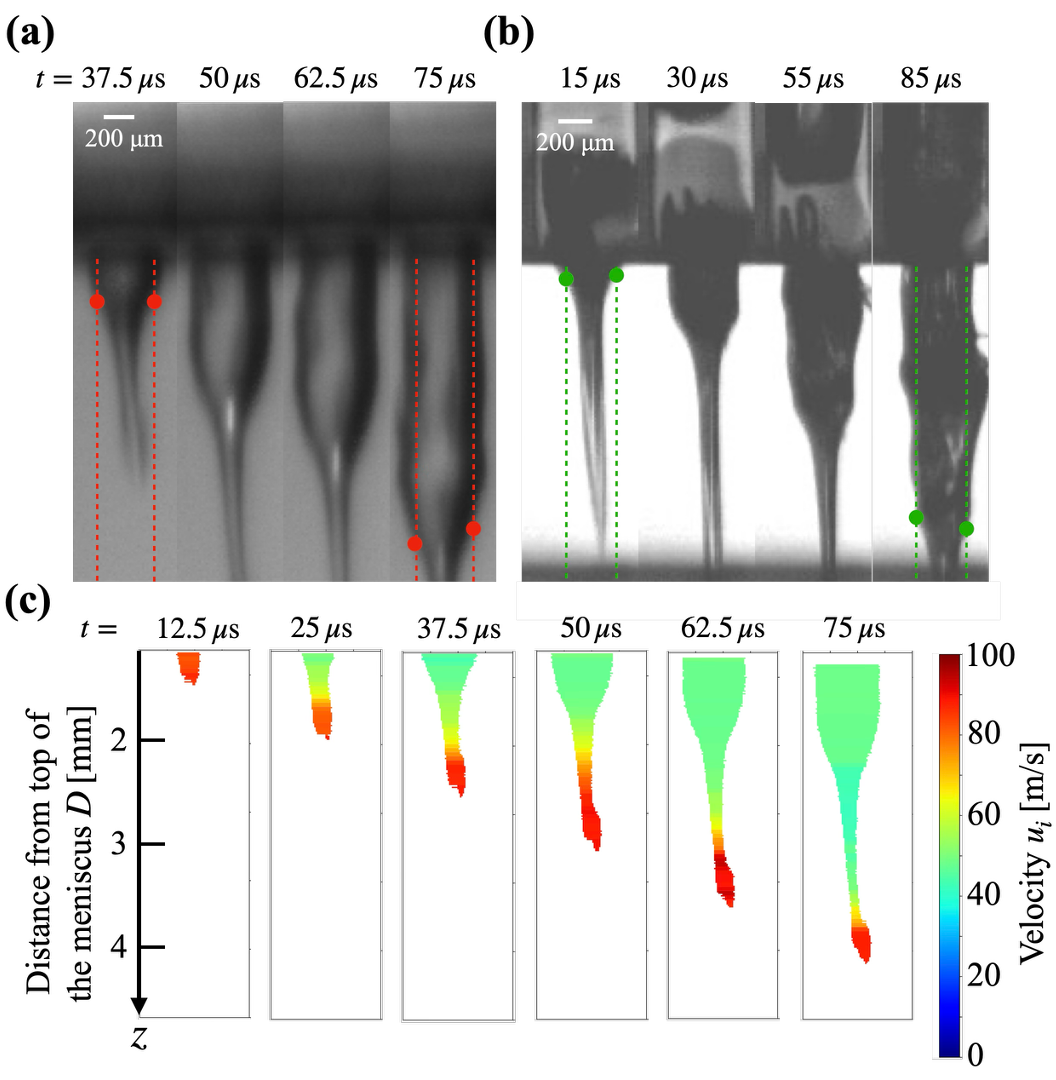}
        \caption{A close-up of the experimental image sequence of the ejection before penetration by (a) an impact-induced jet with \(V_{\mathrm{jet\,root}} = 49.3~\mathrm{m/s}\)\,(\(V_{\mathrm{jet\,tip}} = 88.8~\mathrm{m/s}\)) and (b) a laser-induced jet with \(V_{\mathrm{jet\,root}} = 23.8~\mathrm{m/s}\) (\(V_{\mathrm{jet\,tip}} = 86.8~\mathrm{m/s}\)). (c) Velocity distribution maps within the impact-induced liquid jet at different times from $t = 12.5~\mu$s to $75~\mu$s, illustrating the temporal evolution of the internal velocity field as the liquid jet is ejected and elongated.   }
        \label{fig:Velocity_distribution_result}
    \end{center}
\end{figure}

From Fig.~\ref{fig:Velocity_distribution_result}(a)(b), which shows the ejection images of the impact-induced jet and the laser-induced jet with the same jet tip velocities as those in Fig.~\ref{fig:Penetration_behavior_image_1}(a)(b), we find that the velocity at the root of the impact-induced jet is \(V_{\text{jet\,root}} = 49.3~\mathrm{m/s}\), while that of the laser-induced jet is \(V_{\text{jet\,root}} = 23.8~\mathrm{m/s}\). Despite this difference, the jet tip velocities are almost the same for the impact-induced (\(V_{\text{jet\,tip}} = 88.8~\mathrm{m/s}\)) and laser-induced jets (\(V_{\text{jet\,tip}} = 86.8~\mathrm{m/s}\)). Here, \(V_{\text{jet\,root}}\) was evaluated as the average velocity calculated from the displacements of two points located symmetrically at a distance of 150~\(\mu\mathrm{m}\) from the nozzle center. These findings indicate that the root velocity in the impact-induced jet is higher than that in the laser-induced jet, and is also greater than the $15$--$20~\mathrm{m/s}$ reported in a previous study~\citep{tagawa2013needle}, which was identified as the penetration velocity required for $5~\mathrm{wt}\%$ gelatin.

\begin{figure}[htbp]
    \begin{center}
        \includegraphics[width=1\columnwidth]{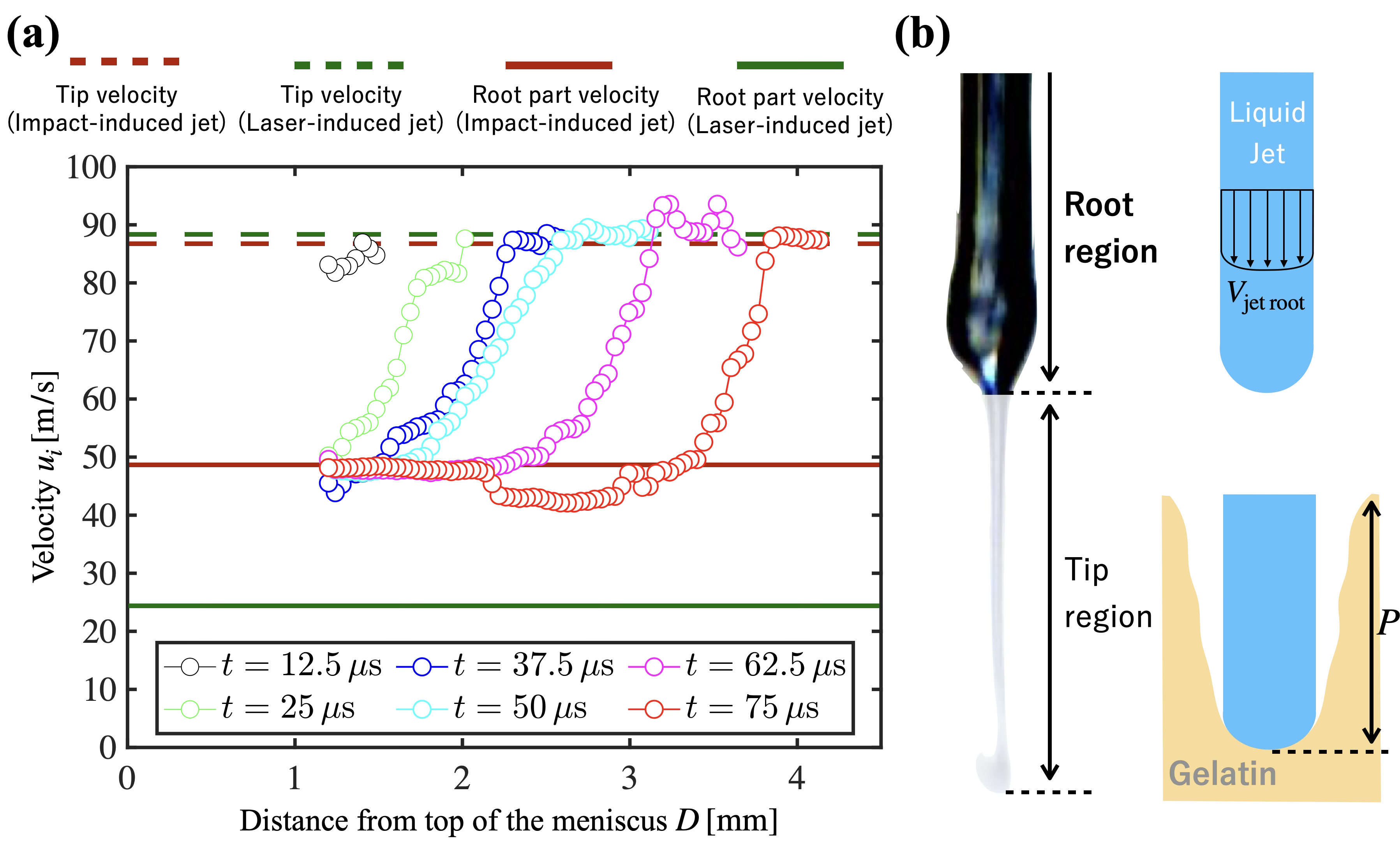}
        \caption{(a)
Velocity distributions of the impact-induced jet along the distance from the top of the initial meniscus. 
Open circles represent the measured axial velocity $u_i$ of the impact-induced jet at different times ($t = 12.5$--$75~\mu\mathrm{s}$), showing the temporal evolution of the jet tip moving downstream. 
The dashed lines represent the tip velocities of the impact-induced (\(V_{\mathrm{jet\,tip}} = 86.8~\mathrm{m/s}\)) and laser-induced jets (\(V_{\mathrm{jet\,tip}} = 88.8~\mathrm{m/s}\)), while the solid lines indicate the root-region velocities of the impact-induced (\(V_{\mathrm{jet\,root}} = 49.3~\mathrm{m/s}\)) and laser-induced jets (\(V_{\mathrm{jet\,root}} = 23.8~\mathrm{m/s}\)). 
This comparison clarifies the differences in acceleration behavior between the two jet generation mechanisms. (b) A high-speed image and a schematic of an impact-induced liquid jet defining the tip and root regions. The root region, which has a larger jet diameter, penetrates into the gelatin and mainly contributes to penetration. The penetration depth $P$ is defined based on the velocity of the root region \(V_{\mathrm{jet\,root}}\) of the jet.
}
        \label{fig:Velocity_distribution_graph}
    \end{center}
\end{figure}

To obtain an accurate velocity profile within the impact-induced jet, we investigate this root velocity in more detail by calculating the velocity distribution within the impact-induced jet. We calculate the velocity distribution in the \(z\)-direction of the impact-induced jet. Using the velocity distribution measurement method reported by previous studies~\citep{igarashi2024optimal, van2014velocity}, the impact-induced jet was binarized, and the velocity distribution within the jet was calculated. The detailed methodology used to compute the velocity distribution is described in \ref{appendix:velocity_analysis}. From Fig.~\ref{fig:Velocity_distribution_result}(c), it is confirmed that, similarly to the laser-induced jet reported by the previous study\,\cite{igarashi2024optimal}, the tip velocity of the liquid jet was the highest and gradually decreased toward the root region of the jet. Figure~\ref{fig:Velocity_distribution_graph}(a) shows the axial velocity $u_i$ within the impact-induced jet as a function of the distance $D$. The velocity distribution reveals two characteristic regions: a high-velocity region near the jet tip and a root region with an almost constant velocity. In Fig.~\ref{fig:Velocity_distribution_graph}(a), the jet tip velocities of the laser-induced jet ($V_{\text{jet\,tip}} = 88.8~\mathrm{m/s}$) and the impact-induced jet ($V_{\text{jet\,tip}} = 86.8~\mathrm{m/s}$) are indicated by dashed lines, while the root velocities for the laser-induced jet ($V_{\text{jet\,root}} = 23.8~\mathrm{m/s}$) and the impact-induced jet ($V_{\text{jet\,root}} = 49.3~\mathrm{m/s}$) are indicated by solid lines (laser-induced jet: green, impact-induced jet: red). These results reveal that, while the jet tip velocities are similar between the impact-induced jet and the laser-induced jet, the jet root velocity in the impact-induced jet is approximately twice as large and remains nearly constant regardless of the distance $D$. This higher, distance-independent root velocity plays a key role in the larger penetration depth.

These results suggest that the higher velocity at the root of the impact-induced jet, compared to that of the laser-induced jet, plays a significant role in producing a larger penetration depth. The recent study \citep{igarashi2024optimal} reported that when the jet can be approximated as having a cylindrical shape, the penetration depth becomes insensitive to the offset distance and is primarily governed by the momentum-related characteristics of the jet, which scale with the liquid density and the square of the jet velocity, $\rho v^{2}$. Consistent with this framework, the present experiments show that the root region of the impact-induced jet can be regarded as cylindrical, as shown in Fig.~\ref{fig:Velocity_distribution_graph}(b). Consequently, penetration dominated by the root region leads to a penetration depth that remains nearly constant regardless of the offset distance $D$ between the top of the meniscus inside the nozzle and the gelatin surface. This behavior stands in contrast to laser-induced jets, for which penetration is governed by a time-evolving conical, focused jet tip and is sensitive to the offset distance.

\clearpage
\section{Penetration model development}
In the penetration of soft materials by liquid jets, penetration occurs only when the jet velocity exceeds a certain threshold known as the critical penetration velocity $U_{\mathrm{c}}$~\citep{tagawa2013needle, uth2013unsteady}. The critical penetration velocity $U_{\mathrm{c}}$ is defined as the minimum jet velocity required to initiate penetration into the target material. Below this threshold ($U_{\mathrm{jet}} < U_{\mathrm{c}}$), the jet only compresses the surface of the material, whereas for $U_{\mathrm{jet}} \geq U_{\mathrm{c}}$, shear deformation and actual penetration occur.
Previous studies have shown that $U_{\mathrm{c}}$ is governed by a balance between the inertial force of the liquid jet and the elasticity of the target material. In particular, the Weber number, defined as
$We = {\rho U_{\mathrm{c}}^{2} d_{\mathrm{jet}}}/{\sigma}$,
follows an empirical relationship with the shear modulus $G$~\citep{van2023microfluidic}:

\begin{eqnarray}
\label{We}
We &\propto& G^{0.83}.
\end{eqnarray}

\noindent
Therefore, the critical penetration velocity is expressed as

\begin{eqnarray}
\label{Uc}
U_{\mathrm{c}} &\propto& G^{0.415}
\sqrt{\frac{\sigma}{\rho d_{\mathrm{jet}}}}.
\end{eqnarray}

\noindent
In this study, the proportionality constant was estimated from the experimental data, and the critical penetration velocity was determined as
$U_{\mathrm{c}} = 2.77\, G^{0.415} \sqrt{\sigma / (\rho d_{\mathrm{jet}})}$.

\begin{figure}[htpb] 
\begin{center} 
\includegraphics[width=1\columnwidth]{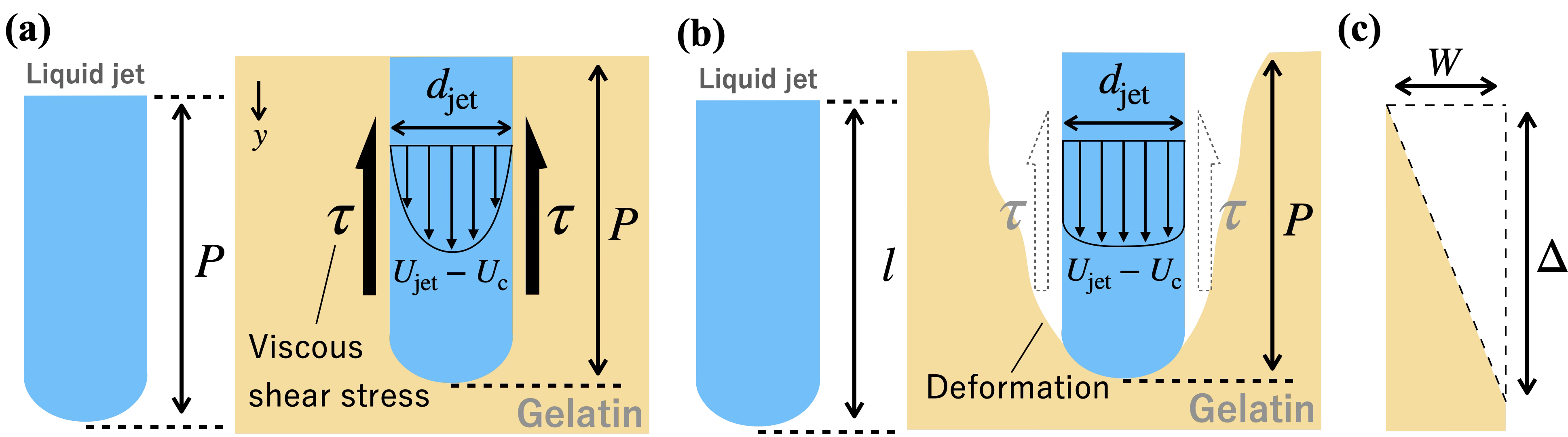} 
\caption{(a) An illustration of the viscous shear stress $\tau$ acting on the side surface of the jet, where the jet (indicated in blue) penetrates into the gelatin. In the viscous shear stress model, the length of the liquid jet inside the target is assumed to be approximately equal to the penetration depth $P$. 
(b) An illustration of the shear deformation of the gelatin as an alternative representation of the viscous shear stress $\tau$. In the shear deformation model, the effective jet length is defined as $l\neq P$. 
(c) A depiction of the shear deformation, where $W$ denotes the cavity width in the material and $\Delta$ represents the cavity depth of the material induced by the applied force.
}
\label{fig:Penetration_model} 
\end{center} 
\end{figure}

In this section, the experimental results are analyzed using the viscous shear stress model. A focused liquid jet penetration model proposed in a previous study~\citep{tagawa2013needle} is adopted. A schematic of the viscous shear stress model is shown in Fig.~\ref{fig:Penetration_model}(a). In this model, during penetration into the skin-simulating material, the flow inside the liquid jet is assumed to be a Hagen--Poiseuille flow~\citep{miyazaki2021dynamic, tagawa2013needle, chung2025computational}. The velocity gradient generates a viscous shear stress $\tau$ acting on the side surface of the penetrating jet, and the work done by this viscous shear stress dissipates the kinetic energy of the jet. Accordingly, the penetration depth $P$ can be expressed as a function of the effective jet velocity $U_{\mathrm{jet}} - U_{\mathrm{c}}$ and a fitting parameter $c_v$ as

\begin{eqnarray}
\label{PM}
\frac{P}{d_{\rm{jet}}} &=& \frac{U_{\rm{jet}}-U_{\rm{c}}}{c_v}.
\end{eqnarray}
\noindent
Based on the results of the previous section, which showed that penetration of the impact-induced jet is governed by the cylindrical root region of the jet, we define the effective jet velocity as $V_{\mathrm{jet\,root}} - U_{\mathrm{c}}$.

\begin{figure}[htbp]
    \begin{center}
        \includegraphics[width=1\columnwidth]{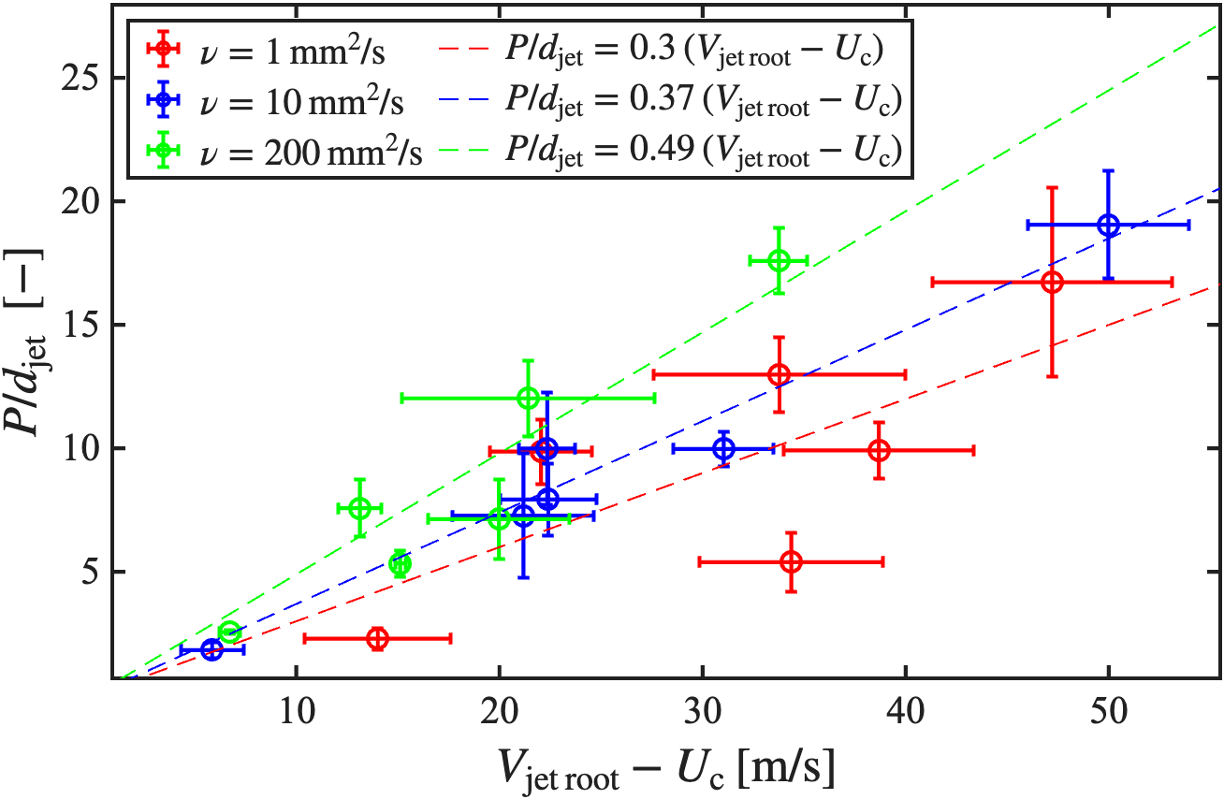}
        \caption{The normalized penetration depth $P/d_{\mathrm{jet}}$ as a function of the relative jet velocity $(V_{\mathrm{jet\,root}}-U_c)$ for three kinematic viscosities: $\nu=1$~mm$^2$/s (red), $\nu=10$~mm$^2$/s (blue), and $\nu=200$~mm$^2$/s (green). Symbols with error bars denote the experimental measurements, with error bars indicating $\pm 1$ standard deviation, while dashed lines indicate the corresponding linear fits: $P/d_{\mathrm{jet}}=0.30\,(V_{\mathrm{jet\,root}}-U_c)$, $P/d_{\mathrm{jet}}=0.37\,(V_{\mathrm{jet\,root}}-U_c)$, and $P/d_{\mathrm{jet}}=0.49\,(V_{\mathrm{jet\,root}}-U_c)$.}
        \label{fig:P_Vjet}
    \end{center}
\end{figure}

Figure~\ref{fig:P_Vjet} shows the relationship between the effective jet velocity $V_{\mathrm{jet\,root}}-U_c$ and the normalized penetration depth $P/d_{\mathrm{jet}}$. As indicated by Eq.~(\ref{PM}), a linear relationship between $P/d_{\mathrm{jet}}$ and $V_{\mathrm{jet\,root}}-U_{\mathrm{c}}$ is observed for each kinematic viscosity. Here, we focus on the physical meaning of the fitting parameter $c_v$, as derived from the viscous shear stress $\tau$ acting on the side surface of the penetrating jet in Fig.~\ref{fig:Penetration_model}(a). Assuming that the internal flow of the jet can be approximated by a Hagen--Poiseuille profile, the viscous shear stress at the jet surface scales as $\tau \sim \mu U_{\mathrm{jet}}/d_{\mathrm{jet}}$, where $\mu$ is the dynamic viscosity. During penetration, this viscous shear stress acts on the lateral surface of the jet along the length of the penetration $P$. Therefore, the total viscous stress work $W_{\mathrm{\tau}}$ is obtained by integrating $\tau$ over the side area $\pi d_{\mathrm{jet}} P$, leading to a characteristic viscous contribution proportional to $\mu U_{\mathrm{jet}} P^{2}$. By balancing this viscous dissipation due to the viscous shear stress $\tau$ with the kinetic energy of a liquid jet segment whose length is equal to the penetration depth $P$, the fitting parameter $c_v$ is physically expressed as a coefficient representing the strength of viscous energy dissipation during penetration. As shown in detail in \ref{appendix:viscous_shear}, where the full derivation is provided, this energy balance yields

\begin{eqnarray}
\label{c_v}
c_v \propto \frac{\nu}{d_{\mathrm{jet}}}.
\end{eqnarray}

\noindent
Therefore, $c_v$ physically represents the ratio between the characteristic jet length scale and the viscous scale of the liquid.
Since $d_{\mathrm{jet}}$ is constant in our experiments, $c_v$ is expected to vary significantly with the kinematic viscosity according to Eq.~\eqref{c_v}. However, such a dependence is not clearly observed in Fig.~\ref{fig:P_Vjet}. In the figure, linear fits are applied to the data for each kinematic viscosity, and the inverse of the slope of each fitted line corresponds to the fitting parameter $c_v$ in Eq.~\eqref{PM}. Although $c_v$ should be proportional to the kinematic viscosity $\nu$, the obtained values do not exhibit a clear monotonic dependence on $\nu$.
This can be explained by the fact that a cylindrical liquid jet penetrating into gelatin does not maintain continuous contact with the target along its side surface, as shown in Fig.~\ref{fig:Penetration_model}(b). Recent studies on liquid jet penetration models have also reported voids along the sides of the cylindrical jets due to backflow on penetration~\citep{anderson2017analytical, uth2013unsteady, babaiasl2020predictive}. Such backflow prevents the generation of viscous shear stress at the jet--gelatin interface. As a result, viscous shear dissipation along the jet side is negligible, which explains why $c_v$ does not show a clear dependence on the kinematic viscosity.

We propose a new penetration model, the shear deformation model, that accounts for voids between the gelatin and the liquid jet. A schematic diagram of the shear deformation model is shown in Fig.~\ref{fig:Penetration_model}(b). Due to the formation of these voids, viscous shear stress does not occur on the side of the jet, and the kinetic energy of the jet is dissipated through compressive deformation and shear deformation. In this model, the shear deformation energy density $e_{\rm{s}}$, defined as the energy required per unit volume for gelatin to undergo shear deformation, is expressed by Eq.~\eqref{es}.

\begin{align}
\label{es}
e_{\rm{s}}&=\frac{ G  {\gamma}^{2}}{2}
\end{align}

\noindent{Considering the shear deformation shown in Fig.~\ref{fig:Penetration_model}(c), we assume for simplicity that the diameter of the conical penetration cavity is of the same order as $d_{\rm jet}$. Consequently, we set the width of the penetration cavity $W=d_{\rm jet}/2$ and the depth of the penetration cavity $\Delta=P$ in this study, and assume the shear strain $\gamma \approx 2P/d_{\mathrm{jet}}$ and the volume of conical penetration $v_p \approx \pi P d_{\mathrm{jet}}^2 / 12$. Under these assumptions, the total shear deformation energy $E_{\rm s}$ consumed by penetration of the liquid jet can be expressed as Eq.~\eqref{Es}.

}

\begin{align}
\label{Es}
E_{\rm{s}} &= e_{\rm{s}} \cdot v_p \notag \\
  &= \frac{\pi G P^{3}}{6}
\end{align}

\noindent
The jet length is approximated by the penetration depth $P$ within the framework of the viscous shear stress model~\citep{tagawa2013needle}, as shown in Fig.~\ref{fig:Penetration_model}(a). The jet length, denoted by $l$, is treated as an independent parameter in the shear deformation model, as shown in Fig.~\ref{fig:Penetration_model}(b). This difference is made because the penetration depth $P$ varies depending on the stiffness of the target material, 
whereas the liquid jet is generated under fixed conditions. 
If the jet length were assumed to vary with the penetration depth $P$,
it would lead to a physically unrealistic situation 
where the same jet would appear to have different lengths depending on the stiffness of the materials. 
Therefore, the jet length $l$ is defined independently of $P$ to describe the intrinsic characteristics of the jet itself. Based on this definition, the kinetic energy of the liquid jet, \( E_{\mathrm{jet}} \), can be formulated as:

\begin{align}
\label{Ejet_2}
E_{\rm{jet}}&= \frac{1}{2} m_{\rm{jet}} (U_{\rm{jet}}-U_{\rm{c}})^2 \notag \\
                 &= \frac{1}{8} \pi \rho d_{\rm{jet}}^{2} l (U_{\rm{jet}}-U_{\rm{c}})^2
\end{align}

\noindent{The fracture of soft materials mainly occurs through crack propagation accompanied by elastic and viscoelastic deformation~\citep{creton2016fracture, duncan2020cutting}. Thus, under the present conditions, energy dissipation due to jet-induced shear deformation is considered to play the primary role. If the kinetic energy of the liquid jet is entirely converted into shear deformation energy of the gelatin, i.e., $E_{\mathrm{s}} = E_{\mathrm{jet}}$, the dimensionless penetration depth $P / d_{\mathrm{jet}}$ can be expressed by Eq.~\eqref{P_djet}, based on Eqs.~\eqref{Es} and~\eqref{Ejet_2}.

}

\begin{eqnarray}
\label{P_djet}
\frac{P}{d_{\mathrm{jet}}} =0.91\cdot\left(\frac{\rho (U_{\rm{jet}}-U_{\rm{c}})^2}{G}\cdot \frac{l}{d_{\mathrm{jet}}} \right)^{\frac{1}{3}}
\end{eqnarray}

\noindent{By introducing the root velocity of the liquid jet $V_{\mathrm{jet\,root}}$, the final expression for the dimensionless penetration depth $P / d_{\mathrm{jet}}$ is written as}

\begin{eqnarray}
\label{P_djet_fin_1}
\frac{P}{d_{\mathrm{jet}}} = 0.91\cdot\left(\frac{\rho (V_{\mathrm{jet\,root}} - U_{\mathrm{c}})^{2}}{G}\cdot\frac{l}{d_{\mathrm{jet}}}\right)^{\frac{1}{3}}.
\end{eqnarray}

\noindent{Regarding the term $l / d_{\mathrm{jet}}$ in Eq.~(\ref{P_djet_fin_1}), assuming that the liquid jet is ejected during the impact duration of the ejector, the jet length $l$ can be expressed in terms of the ejector's impact duration $\Delta t$ and the jet root velocity $V_{\mathrm{jet\,root}}$ as}
\begin{eqnarray}
\label{l}
l \sim \Delta t \cdot V_{\mathrm{jet\,root}}.
\end{eqnarray}

\noindent{The duration of impact $\Delta t$ has been reported in previous studies to be approximately $\Delta t = 1.2 \times 10^{-4} \, [\mathrm{s}]$ for collisions with metal substrates~\cite{kurihara2024pressure}. In the present study, the impact duration between the ejector and the metal block was also assumed to be approximately $1.2 \times 10^{-4} \, [\mathrm{s}]$. 
}

\begin{figure}[htbp]
    \begin{center}
        \includegraphics[width=1\columnwidth]{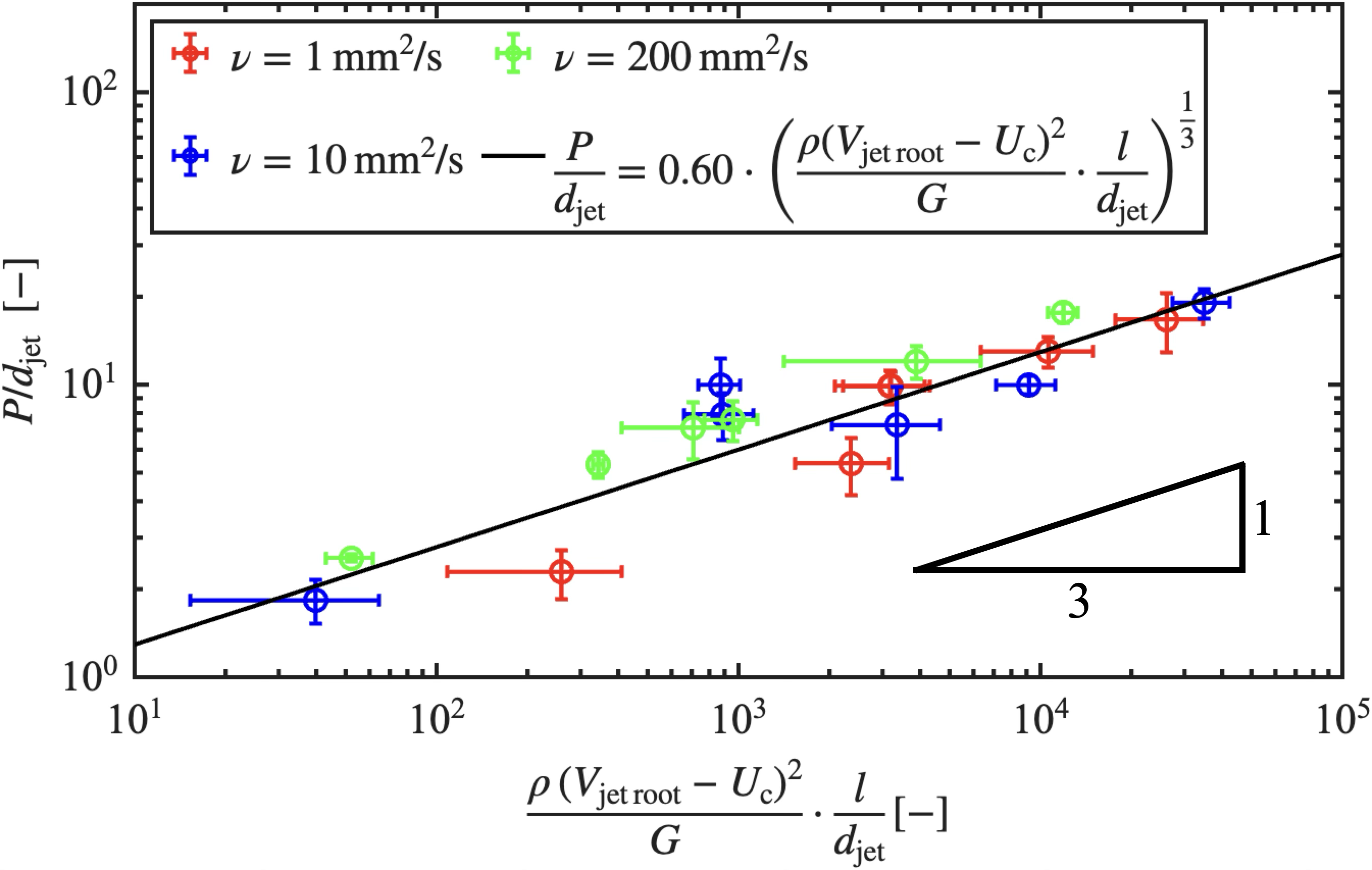}
        \caption{Experimental results for the dimensionless penetration depth \( P/d_{\mathrm{jet}} \)
plotted against the dimensionless parameter 
\( \rho (V_{\mathrm{jet\,root}} - U_{\mathrm{c}})^2 / G \cdot (l / d_{\mathrm{jet}}) \),
together with the fitting curve based on the shear deformation model expressed in Eq.~\eqref{P_djet_fin_1}.
Symbols denote the experimental measurements, with error bars indicating \(\pm 1\) standard deviation.
The solid line represents the best-fit result with a fitting coefficient of 0.60,
showing good agreement between the experimental data and the prediction of the shear deformation model.
}
        \label{fig:Final_result}
    \end{center}
\end{figure}

The experimental results are shown in Fig.~\ref{fig:Final_result}, with the horizontal axis representing $\rho (V_{\mathrm{jet\,root}} - U_{\mathrm{c}})^{2}/G \cdot (l/d_{\mathrm{jet}})$ and the vertical axis representing $P/d_{\mathrm{jet}}$. As seen in Fig.~\ref{fig:Final_result}, the experimental results are consistent with the shear deformation model described by Eq.~\eqref{P_djet_fin_1}, capturing the scaling trend associated with the exponent in the model well. The fitting coefficient obtained for the shear deformation model from the experimental data was 0.60. This coefficient of 0.60, considering the coefficient 0.91 in Eq.~\eqref{P_djet_fin_1}, indicates that approximately 66\% of the jet’s kinetic energy is effectively converted into shear deformation energy within the gelatin, while the remaining energy is likely dissipated through viscous losses and local deformations of the material. Such a magnitude of deviation from unity is reasonable considering the simplifications introduced in the shear deformation model, including the assumptions of cylindrical jet geometry and uniform shear deformation, and thus the result provides quantitative support for the proposed shear deformation mechanism.

These findings confirm the validity of the shear deformation model and suggest that it serves as a viable means of describing the liquid jet penetration process. Furthermore, although numerous theoretical frameworks such as the viscous shear stress model~\citep{tagawa2013needle} and other penetration models~\citep{baxter2005jet,van2023microfluidic, ohl2025penetration, mousavi2025modelling} have
been proposed to describe liquid jet interactions with skin-simulating materials, this study has successfully established a shear deformation model that is
particularly suited to cylindrical jet penetration. Using an
impact-induced jet system capable of ejecting even highly viscous liquids,
we were able to systematically examine the effects of viscosity over a wide range of kinematic viscosities on penetration dynamics and validate the applicability of the proposed model.
This achievement not only provides a robust physical framework for
understanding the liquid jet--substrate interactions, but also offers a new pathway for
controlling penetration behavior. The shear deformation model holds great potential for
future applications in both academic and engineering fields, including the
optimization of needle-free injection systems, the precise processing of microfluidic materials, and the controlled delivery of viscous biofluids in medical and
biomanufacturing technologies.

\clearpage

\section{Conclusions and outlook}
This study has clarified the penetration mechanism of impact-induced jets into skin-simulating materials. Comparative experiments with
laser-induced jets showed that, even at similar jet tip velocities,
impact-induced jets achieve significantly deeper penetration. This behavior originates from the high and nearly uniform velocity of the cylindrical jet root region, which is the primary determinant of the penetration depth. Consequently, the penetration depth remains largely independent of the offset distance, highlighting a fundamental difference from laser-induced jet penetration, in which the penetration depth is governed by changes in the liquid jet tip morphology caused by the offset distance. This demonstrates superior controllability of the impact-induced jet for practical applications.

The systematic variation of jet velocity, liquid viscosity, and material shear modulus revealed clear deviations from the conventional viscous shear stress model. Motivated by these discrepancies, we proposed a shear deformation model in which the jet kinetic energy is mainly dissipated through shear deformation of the material. The model successfully reproduced the experimental results across a wide range of inertial and viscous forces of the liquid jet and shear modulus (elastic modulus) of the material, providing a unified physical framework for liquid jet penetration.

These findings establish, through the use of skin-simulating materials, a foundational physical understanding of liquid jet penetration into soft materials, including biological tissues. In doing so, this study has succeeded in constructing a basic knowledge framework for liquid jet penetration phenomena, providing an essential basis for extending
jet-based delivery technologies to real biological system.
While the present study focuses on single-layer skin-simulating materials,
the proposed framework provides a mechanistic baseline that can be extended
to layered tissues and post-penetration transport processes.
Incorporating epidermal barriers and in vivo tissue heterogeneity represents
a natural next step, guided by the predictive capability established here.

\clearpage

\appendix
\section{Velocity distribution measurement method}
\label{appendix:velocity_analysis}
\begin{figure}[htbp]
    \begin{center}
        \includegraphics[width=1\columnwidth]{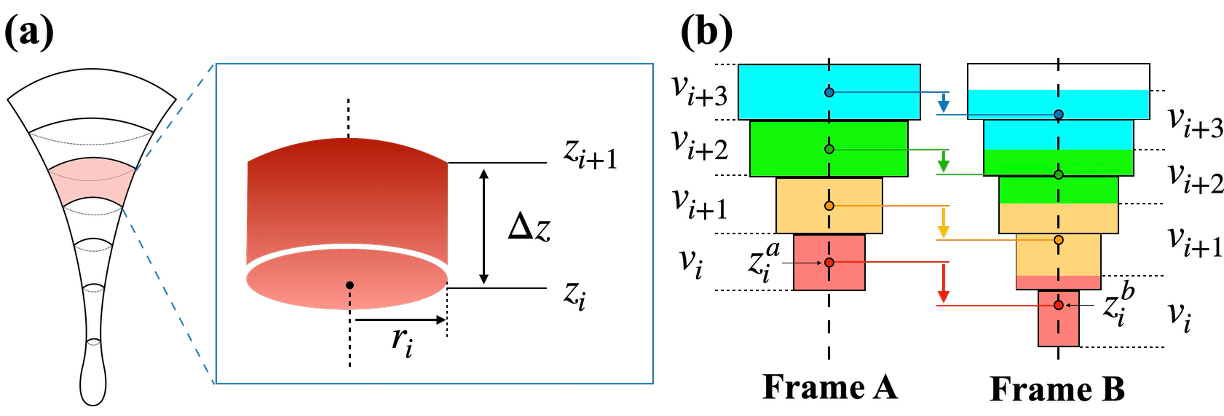}
        \caption{(a) Modeling the liquid jet as a stack of cylinders and dividing it into volume elements, (b) estimating the velocity distribution $u_i$ using these volume elements, and (c) binarized liquid jet images and color maps of the velocity distributions $u_i$ obtained from the experimental images.}
        \label{fig:Velocity_distribution_method}
    \end{center}
\end{figure}

    We used a method established in a previous study~\citep{van2014velocity}, in which the velocity distribution is estimated by calculating the displacement of the centroid of each volume element within the liquid jet. The liquid jet is assumed to be an incompressible and axisymmetric flow.
As shown in Fig.~\ref{fig:Velocity_distribution_method}(a), the jet is modeled as a structure composed of stacked cylindrical segments, which are divided into pixel-based volume elements along the \(z\)-direction. The volume \(v_i\) of each element is estimated using its radius \(r_i\) and height \(\Delta z\), as given in Eq.~\eqref{Volume_jet_velocity}.
\begin{eqnarray}
\label{Volume_jet_velocity}
v_i=\pi r_i^2 \Delta z
\end{eqnarray}

\noindent{The velocity in the $z$-direction can be obtained by tracking the displacement of the centroid of each volume element. Figure~\ref{fig:Velocity_distribution_method}(b) compares frames A and B, separated by a time interval $\Delta t$. For the same volume element, the centroids ${z_{i}}^{a}$ in frame A and ${z_{i}}^{b}$ in frame B are calculated. From these centroid positions, the centroid velocity $u_i$ can be expressed as shown in Eq.~\eqref{centroid_jet_velocity}.
}
\begin{eqnarray}
\label{centroid_jet_velocity}
u_i=\frac{{z_i^b}-{z_i^a}}{\Delta t}
\end{eqnarray}

\label{app1}

\section{Extension of the viscous shear stress model}
\label{appendix:viscous_shear}
Assuming a Hagen--Poiseuille flow profile, the velocity distribution is given by
\begin{equation}
u(r) = u_{\max}\left(1 - \frac{r^2}{R^2}\right),
\end{equation}
where $R = d_{\mathrm{jet}}/2$ is the jet radius.
Therefore, the velocity gradient at the wall (jet surface) is
\begin{equation}
\left. \frac{\partial u}{\partial r} \right|_{r=R} = -\frac{2u_{\max}}{R}
= -\frac{4u_{\max}}{d_{\mathrm{jet}}}.
\end{equation}
The viscous shear stress $\tau$ shown in Fig.~\ref{fig:Penetration_model}(a) is then expressed as
\begin{equation}
\tau = \mu \left. \frac{\partial u}{\partial r} \right|_{r=R}
     = -\frac{2\mu u_{\max}}{R}
     = -\frac{4\mu u_{\max}}{d_{\mathrm{jet}}},
\end{equation}

\begin{eqnarray}
\tau = \frac{4 \mu (U_{\mathrm{jet}}-U_{\mathrm{c}})}{d_{\mathrm{jet}}}.
\end{eqnarray}

\noindent
As shown in Fig.~\ref{fig:Penetration_model}(a), the work $W_{\tau}$ done by the viscous shear stress $\tau$ acting on the side surface of the penetrating jet 
is expressed as

\begin{eqnarray}
\label{W}
W_\tau &=& \int_0^P \tau \, (\pi d_{\mathrm{jet}} y) \, dy \nonumber \\
       &=& \frac{1}{2} \pi \tau d_{\mathrm{jet}} P^2 .
\end{eqnarray}Here, the length of the liquid jet inside the target is assumed to be approximately equal to the penetration depth $P$, 
because the jet maintains a nearly constant shape during penetration and the entire jet body is contained within the target at the maximum penetration moment~\citep{tagawa2013needle}. Substituting this expression for $\tau$ into Eq.~\eqref{W}, 
the total viscous work $W_{\tau}$ is finally expressed as

\begin{eqnarray}
\label{W_fin_2}
W_\tau &=& 2\pi \mu (U_{\mathrm{jet}}-U_{\mathrm{c}}) P^2 
\end{eqnarray}

\noindent
Since the mean velocity of a Hagen--Poiseuille flow is half of the velocity $U_{\mathrm{jet}}-U_{\mathrm{c}}$ ($U_{\mathrm{ave}} = (U_{\mathrm{jet}}-U_{\mathrm{c}}) / 2$), 
the kinetic energy $K_{\mathrm{jet}}$ of the penetrating jet is given by

\begin{eqnarray}
\label{Kjet_2}
K_{\mathrm{jet}} &=& \frac{1}{2} m_{\mathrm{jet}} U_{\mathrm{ave}}^2 \nonumber \\
                 &=& \frac{1}{2} \left( \frac{\pi}{4} \rho d_{\mathrm{jet}}^2 P \right) 
                     \left( \frac{U_{\mathrm{jet}}-U_{\mathrm{c}}}{2} \right)^2 \nonumber \\
                 &=& \frac{1}{32} \pi \rho d_{\mathrm{jet}}^2 P (U_{\mathrm{jet}}-U_{\mathrm{c}})^2 
\end{eqnarray}

\noindent{where $m_{\mathrm{jet}}$ denotes the mass of the penetrating liquid jet and $\rho$ is the density of the liquid. The kinetic energy of the liquid jet before penetration \( K_{\mathrm{jet}} \) and the total viscous shear stress work \( W_{\tau}\) are expressed  as shown in Eq.~\eqref{Kjet_2} and Eq.~\eqref{W_fin_2}.}

\noindent{In the liquid jet penetration process, it is assumed that the kinetic energy \( K_{\mathrm{jet}} \) of the liquid jet just before penetration is dissipated through the energy $W_{\tau}$ caused by the viscous shear stress $\tau$ acting on the side surface of the jet. From this assumption, it follows that $K_{\mathrm{jet}}= W_{\tau}$, which leads to Eq.~\eqref{Re}.}

\begin{eqnarray}
\label{Re}
\frac{P}{d_{\rm{jet}}} &=& \frac{d_{\rm{jet}} (U_{\rm{jet}} - U_{\rm{c}})}{64 \nu} \nonumber \\
                       &=& \frac{Re}{64}
\end{eqnarray}

\noindent{As discussed in Section~3.2, the velocity distribution inside the impact-induced jet reveals that the penetration behavior is dominated not by the jet tip but by the cylindrical root region that continuously interacts with the gelatin during penetration. 
Therefore, the root velocity \(V_{\mathrm{jet\,root}}\) was adopted as the representative jet velocity in the following analysis. The experimental results are plotted with the Reynolds number defined as \( Re = (d_{\mathrm{jet}}/\nu)(V_{\mathrm{jet\,root}} - U_{\mathrm{c}}) \) on the horizontal axis and the normalized penetration depth \( P / d_{\mathrm{jet}} \) on the vertical axis, as shown in Fig.~\ref{fig:Re_result}. However, although each viscosity condition individually shows a proportional relationship between 
the normalized penetration depth \( P / d_{\mathrm{jet}} \) and the Reynolds number \(Re\), 
the proportionality coefficient varies with viscosity. 
Although we attempted to account for the effect of liquid viscosity by using the Reynolds number \(Re\), the data for different viscosities do not collapse onto a single master curve when plotted against \(Re\).
This suggests that the viscosity influence observed in the experiments is not adequately represented by the \(1/\nu\) dependence embedded in the definition of \(Re\). Therefore, in this study, we consider a new penetration model.
 
\begin{figure}[htbp]
    \begin{center}
        \includegraphics[width=1\columnwidth]{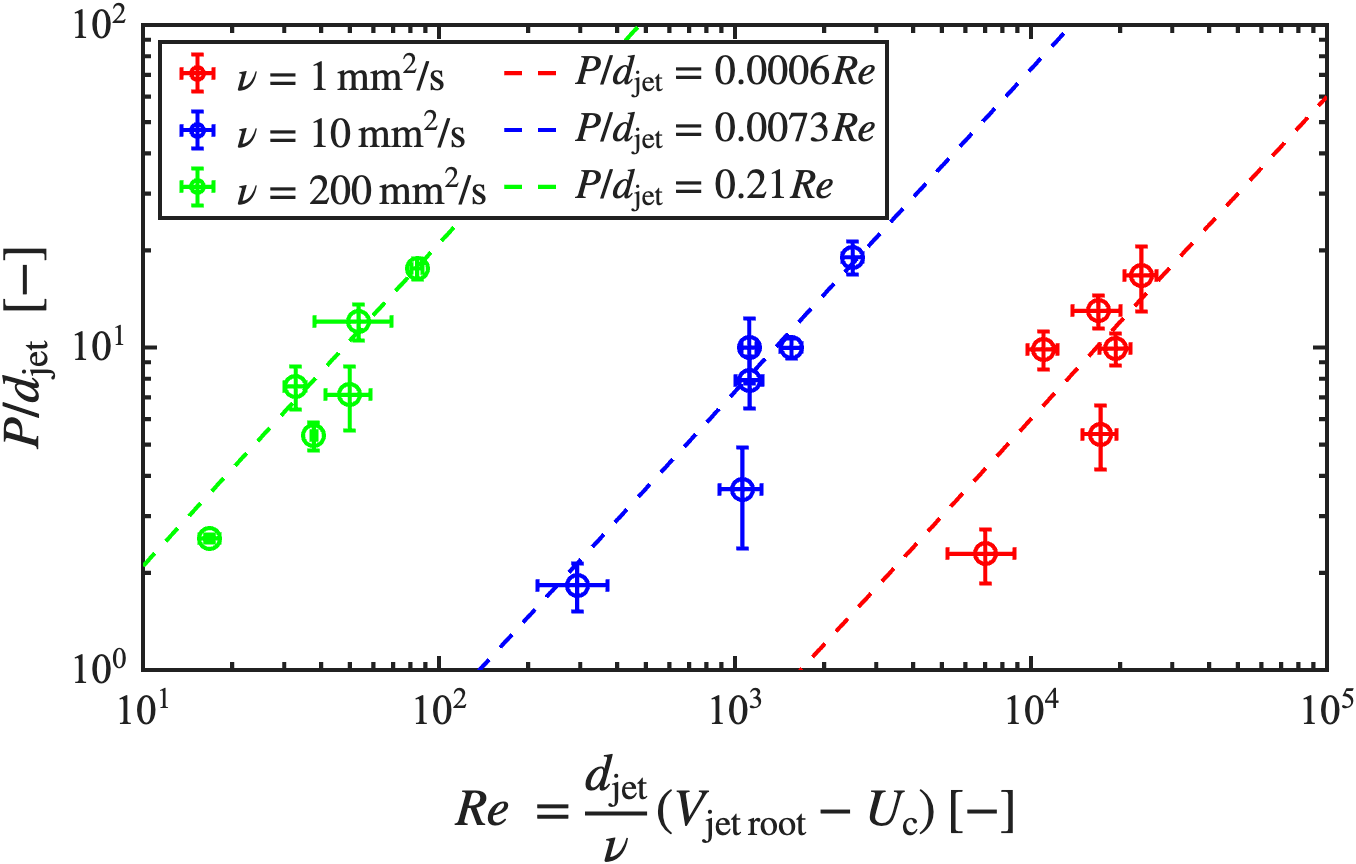}
        \caption{Experimental results of the dimensionless penetration depth $P/d_{\mathrm{jet}}$ as a function of the Reynolds number $Re = (d_{\mathrm{jet}}/\nu)\,(V_{\mathrm{jet\,root}} - U_{\mathrm{c}})$. Each color represents a different kinematic viscosity ($\nu = 1,\, 10,\, 200~\mathrm{mm^2/s}$). Error bars indicate $\pm 1$ standard deviation.}
        \label{fig:Re_result}
    \end{center}
\end{figure}

\section*{Acknowledgments}
This work was funded by the Japan Society for the Promotion of Science (Grant Nos.\ 20H00222, 20H00223, 20K20972, and 24H00289), the Japan Science and Technology Agency (Grant Nos.\ PRESTO JPMJPR2105 and SBIR JPMJST2355) and JST SPRING (Grant No.\ JPMJSP2116).

\section*{Declaration of Interests}
The authors report no conflict of interest.

\bibliographystyle{elsarticle-num}
\bibliography{cas-refs}






\end{document}